# A FAST METHOD FOR IMPLEMENTATION OF THE PROPERTY LISTS IN PROGRAMMING LANGUAGES

Hassan Rashidi[1]

[1] Department of Electrical, Computer and IT Engineering, Islamic Azad University, Qazvin Branch, Iran

`hrashidi@qiau.ac.ir;hrashi@gmail.com`

## ABSTRACT

*One of the major challenges in programming languages is to support different data structures and their variations in both static and dynamic aspects. One of the these data structures is the property list which applications use it as a convenient way to store, organize, and access standard types of data. In this paper, the standards methods for implementation of the Property Lists, including the Static Array, Link List, Hash and Tree are reviewed. Then an efficient method to implement the property list is presented. The experimental results shows that our method is fast compared with the existing methods.*

## 1. INTRODUCTION

Many applications and databases require some mechanisms for storing variable-size data objects of information in some situations [1]. A variable-size data objects is one in which the number of components in an object may change dynamically during program execution. Some of the major types of variable-size data structures are list, list structure, stack, queue, tree, directed graph and property list.

We focus on the property list, which is a list of alternating names and values. As a formal definition for the property list in the standard textbook [1], a record with a varying number of components is termed as property list if the number of components may vary without restriction. In a property list, both the component names (field names) and their values must be stored. Each field name is termed a property name; the corresponding value of the field is the property value. A property list is also a structured data representation used by Cocoa and Core Foundation [2] as a convenient way to store, organize, and access different types of data. The property list is natural to use when the number and type of components in an object are not known in advance. Property List data structure supports many real-time applications when they read some data from an input device or change attributes of objects during program execution.

A common representation for a property list is as an ordinary linked list, with the property names and their values alternating in a single long sequence, as illustrated in Figure 1. In this Figure, the odd number items are property names, and the even are property values. There are three commands to process a property list:

- Inserting a new element to the list: When a new property is inserted in the property list, two components are inserted: the property name and its value.
- Removing an element from the list: To remove a particular property value (e.g., the value for the 'Name' property in Figure 1), the list is searched, looking only at the property names, until the desired property is found. A pair of component is then deleted from the list.

- Finding a value in the list: To select a particular property value (e.g., the value for the 'Age' property in Figure 1), the list is searched, looking only at the property names, until the desired property is found. The next list component is then the value for that property.

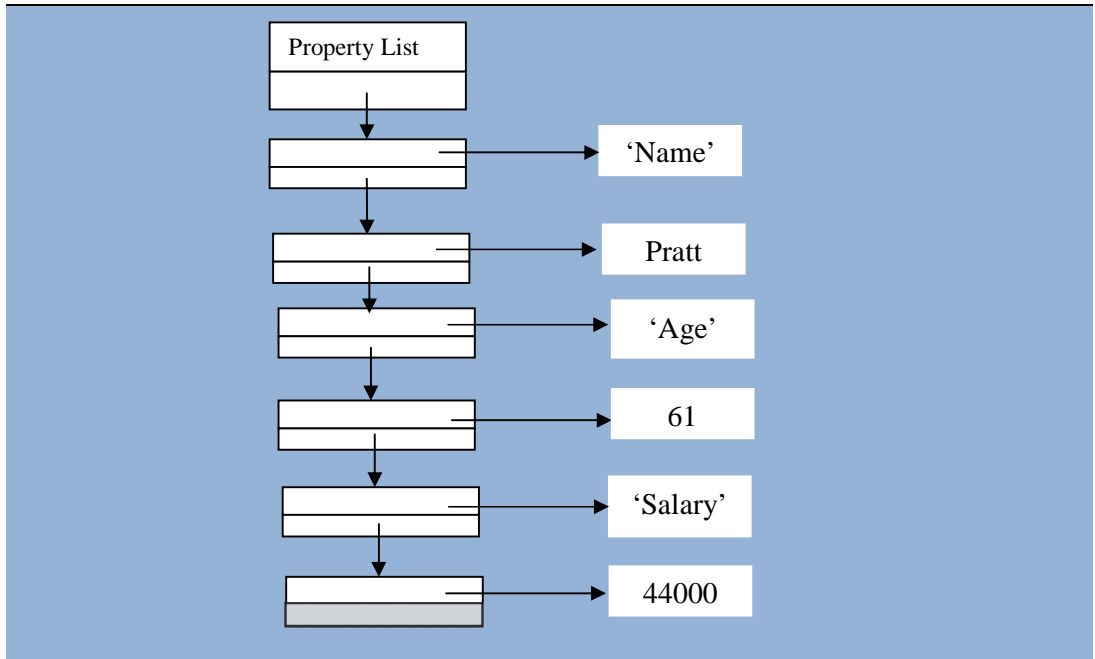

Figure 1. A storage representation of a Property List

The Property lists are used extensively by applications and other software on Mac OS X and iOS [2]. For example, the Mac OS X Finder (through bundles) uses property lists to store file and directory attributes. Applications on iOS use property lists in their Settings bundle to define the list of options displayed to users.

The Property list has a simple XML format, designed by Apple for OSX as a format for storing lists of key-value pairs (see [3]-[4]). In this operation system, most applications store their Preferences as property list files. The property-list programming interfaces for Cocoa and Core Foundation allow the user to convert hierarchically structured combinations of these basic types of objects to and from standard XML. The user can save the XML data to disk and later use it to reconstruct the original objects.

The Property lists are not part of the LISP language, but are an abstraction of common list patterns and usages, and are often defined as LISP library functions. Each item in a list is tagged with a name preceding the item like $( n_1 \ val_1 \ n_2 \ val_2 \ ... \ n_k \ val_k )$. In this list $n_i$ is the property name of i[th] element and $val_i$ is the property value of i[th] element.

This paper presents a fast method for implementation of the Property List in programming languages. Section 2 makes a literature review over the related works in implementation of the property lists and described the four standards methods. Section 3 presents the detail of the method and makes a comparison on the methods in the main features as well as their complexities. Section 4 is dedicated to some experimental results. This section makes a comparison between the times required to perform the operations on the property lists and derives the observations in practise. Section 5 is considered for summary and conclusion.

## 2. THE RELATED WORK

The Property lists organize data into named values and lists of values using several object types. This type of data structure gives the user the means to produce data that is meaningfully structured, transportable, storable, and accessible, but still as efficient as possible. In this section, the standards existing methods for implementation of Property List are reviewed.

### 2.1. Link list Method

The first choice for implementation of property list is link list. For situations where a user needs to store small amounts of persistent data, for example less than a few hundred kilobytes, property lists offer a uniform and convenient means of organizing, storing, and accessing the data. In these situations, the simplest property-list implementation is a linked list (like Figure 1). The users can either have the alternating elements be the keys and values (LISP does this), or they can have each element be a structure containing pointers to the key and value. The linked list implementation is appropriate when the users:

- are just using the pattern to allow user annotations on object instances.
- don't expect many such annotations on any given instance.
- are not incorporating inheritance, serialization or meta-properties into their use of the pattern.

Logically a property list is an unordered set, not a sequential list, but when the set size is small enough a linked list can yield the best performance. The performance of the link list is O(N), so for long property lists the performance can deteriorate rapidly.

If the user needs a way to store large complex graphs of objects, objects not supported by the property-list architecture, or objects whose mutability settings must be retained, use Archiving and Serializations[2]. Archiving and serializations are two ways in which the user can create architecture-independent byte streams of hierarchical data. Byte streams can then be written to a file or transmitted to another process, perhaps over a network. When the byte stream is decoded, the hierarchy is regenerated. Archives provide a detailed record of a collection of interrelated objects and values. Serializations record only the simple hierarchy of property-list values.

### 2.2. Static Array Method

The simplest method to implement the property list is the use of fast static arrays with empty slots. In this method, a large array with a specified size for elements is allocated statically, from beginning to the end of execution. For each component, we have to consider a couple of slots (words). The first one is for the component's name and the second one for component's value. When the software wants to insert a new element, the first empty slot is selected for the position of insertion. So we have to search to find an empty sot. To remove a particular property value (e.g., the value for the 'Name' property in Figure 1), the static array is searched, looking only at the property names, until the desired property is found. After deletion, the pair of slots is marked as the hole/empty slots in the array.

## 2.3. Hash Method

The next most common implementation method is a ***hash-table***, which yields amortized constant-time on the operations of finding, inserting and removing for a given list, albeit at the cost of more memory overhead and a higher fixed per-access cost, i.e. the cost of the hash function. When a confliction occurs on inserting a new element in the list, the solutions of ***Rehashing, Sequential Search*** and ***Bucketing*** [1] may be used.

In most systems, a hash-table imposes too much overhead when objects are expected to have only a handful of properties, up to perhaps two or three dozen [6]. A common solution is to use a hybrid model, in which the property list begins life as a simple array or linked list, and when it crosses some predefined threshold (perhaps 40 to 50 items), the properties are moved into a hash-table. So if we need a tolerant constant-time on access an item and want to maintain the insertion order, we can't do better than a ***_LinkedHashMap_*** [6], a truly wonderful data structure. Java 6.0 implemented this solution. The complexity of this method is O(1). However, the costs of hash-function and its overheads for solving confliction are inevitable.

## 2.4 Binary Tree Method

The 4th method for implementation of property list is binary tree. If a language needs to impose a sort order on property names, it must use an ordered-map implementation, typically an ordered binary tree such as a splay tree or red/black tree (see [5] and [6]). A splay tree can be a good choice because of the low fixed overhead for insertion, lookup and deletion operations, but with the tradeoff that its theoretical worst-case performance is that of a linked list. A splay tree can be especially useful when properties are not always accessed uniformly. If a small subset M of an object's N properties are accessed most often, the amortized performance becomes O(log M), making it a bit like an Least Recently Used (LRU) cache [2].

## 3. OUR METHOD- THE SET

A set is a data object containing and unordered collection of distinct values. In contract, a list is an ordered collection of values, some of which may be repeated. The basic operations on sets are: (a) Search or Lookup a data value in a set, (b) Insertion and Deletion of single values, and (c) Union, Intersection, and Difference of Sets. This section presents a new and fast method for implementation of the property lists.

In some situations, the property-list architecture may prove insufficient [2] and inefficient [9]. Our method for efficient implementation of property list is to present it as a set rather than a list because elements are accessed randomly by subscript (attribute name) rather than sequentially., a root property-list object is at the top of this hierarchy with a couple of pointers like Figure 2.

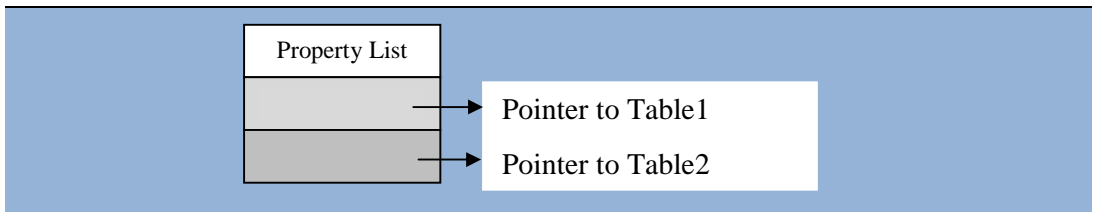

Figure 2. Representation of a property list object

In this method, we define a data structure like Table 1, consisting of ***SetContents*** to store the contents and Pointers to ith element. We make OR operation together all the properties in the set and store this value in ***SetContents***. The number of elements in the property list is variable with a limit on the maximum. For a 32-bit wordsize of memory, there will be 32 such pointers. The names are stored in a global table like Table 2 with a bit value, representing its position in the set. If a machine has a 32-bit word size, then up to 32 properties can be stored, with bit values of: 1, 10, 100, 1000, and so on. The Bit Strings are shown for clarity; i.e. they can be removed practically because they are $i^n$ (n=0, 1, 2,.., 31) respectively. The commands to process the property list in this method are as follows:

- On lookup operation, we check if the bit string of the property name requested ***AND SetContents*** = 0, then the property is not defined in the set. If the value is 1, then the bit position defines the location containing a pointer to its attribute value.
- When a new element is to be inserted in the list, we must make a lookup as above. If the property name is in the list, duplication is not possible. If the result of the lookup is negative, the property name is put into an empty position in the Table 2. Then, we make ***OR*** operation the ***SetContents*** with the corresponding bit string of the property name. After that the corresponding Printer to the property value is set in Table1.
- When an existing element is to be removed from the list, we must make a lookup again. If the result of the lookup is positive, then we make an AND operation of ***SetContents*** with 00000..00 and store the result in ***SetContents***. After that the memory for property name and property vale are freed and they set to ***Null*** in Tables 1 and 2. If the result of the lookup is negative (The result of 0), the element requested doesn't exist in the list.

Table 1. The Data Structure for the method

| SetContents |
| --- |
| **Pointer to Property Value$_1$** |
| **Pointer to Property Value$_2$** |
| ………. |
| ………. |
| **Pointer to Property Value$_{32}$** |

Table 2. A Global Table with a bit string, representing the position of each property value

| Bit String | Pointers to Property Name | |
| --- | --- | --- |
| **00…..01** | | → Property Name$_1$ |
| **00…..10** | | → Property Name$_2$ |
| ……… | | |
| ……… | | |
| **10…..00** | | → Property Name$_{32}$ |

Since the property lists are based on an abstraction for expressing simple hierarchies of data, they can support the application programs. Some types are for primitive values and others are for containers of values. The primitive types are ***strings, numbers, binary data, dates,*** and ***boolean*** values. The containers are ***arrays*** and ***dictionaries.*** The arrays are indexed collections of values and the dictionaries are collections of values each identified by a key. The containers can contain other containers as well as the primitive types. Thus the user might have an array of dictionaries, and each dictionary might contain other arrays and dictionaries, as well as the primitive types. A root property-list object is at the top of this hierarchy, and in almost all cases is a dictionary or an array like Figure 3. Note, however, that a root property-list object does not have to be a dictionary or array; for example, the user could have a single ***string, number,*** or ***date***, and that primitive value by itself can constitute an array or a dictionary.

The Property list is extensively used in Markup Languages. From the basic abstraction, languages derive both a static representation of the property-list data and a dynamic (runtime) representation of the property list  [2]. The static representation of a property list, which is used for storage, can be either XML or binary data. The binary version is a more compact form of the XML property list. In XML, each type is represented by a certain element. The runtime representation of a property list is based on objects corresponding to the abstract types. Both he static and dynamic representation can use our implementation.

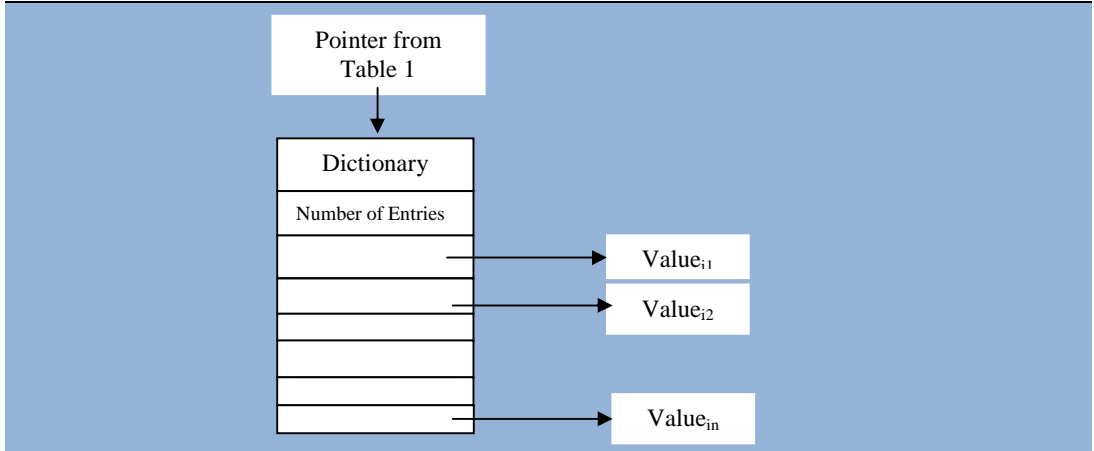

Figure 3. An array or dictionary of collections of values

The property lists are also used to define the SQL Queries in Database Management Systems and JDBC2 Components [8]. The method presented here can by used in these systems and can create more effective Dynamic Data Object.

The performance of this method for all operations is O(1). On insertion and deletion operation of an element, although the performance of this method is as the same as the hash-function, it benefits from the lookup operation and practically outperforms the hash because of no overheads.

Table 3 makes a summary on the main features and performance of the five methods discussed in this paper.  Although the Static Array is simple to implementations, the complexities of the operations are higher. In the Link List method, the property list is presented as an unordered set and has a lower difficulty for implementation. After that, the hash and binary tree are with a high and medium difficulty for implementation, respectively. Both methods have some overheads in run-time during the operations. As shown in the table, our method has no overheads on operations and its implementation is easy.

Table 3. A comparison among the methods for implementation of property lists

| Methods | Main Features | Complexity of Operations | | |
|---|---|---|---|---|
| | | Lookup | Insertion | Deletion |
| Link List | An unordered set, Low difficulty to implement | O(N) | O(1) | O(N) |
| Hash | High fixed overhead, High difficulty to implement | O(1) | O(1) | O(1) |

| Binary Tree | Low fixed overhead, Medium difficulty to implement | O(Log N) | O(Log N) | O(Log N) |
|---|---|---|---|---|
| Static Array | Simple to implement, waste of memory | O(N) | O(N) | O(N) |
| Our Method (Set) | No overhead, Easy to implement | O(1) | O(1) | O(1) |

## 4. SOME EXPERIMENTAL RESULTS

In this section, several experiments are run to compare the standard algorithms and the method presented in this paper. The software was implemented in Borland C++ and then was run to compare the time required to perform several operations on a GenuineIntel 2.599 GHz PC with 1 GMB RAM on Windows XP. Figure 4 shows the main snapshot of our software.

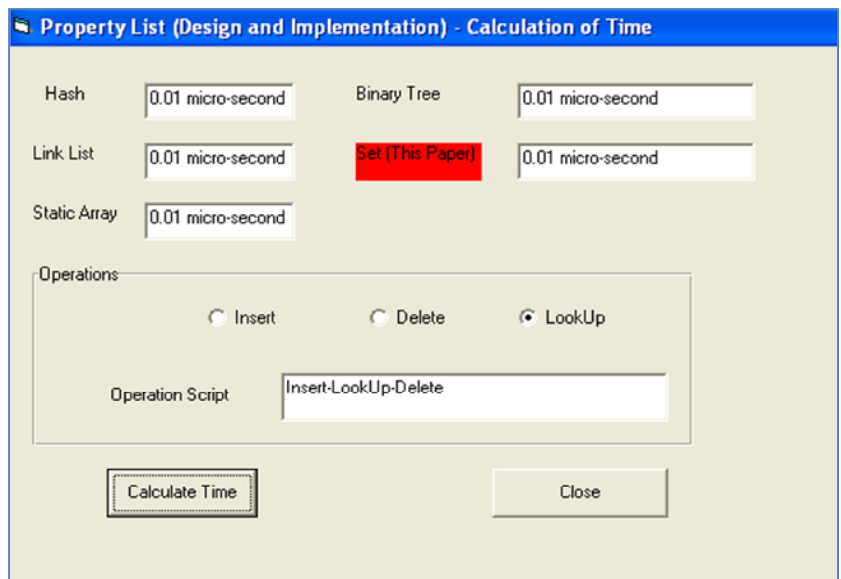

Figure 4. The main snapshot of the software

The maximum size in the property list is assumed large up to 32 components. At the initial time, between 25 and 29 components are inserted into the property list so that we can run three operations in the same kinds. The length of Name component in the property lists is limited to 32 characters. In the three operations (Insert, Delete and LookUp), the position of insertion, the elements to be Deleted and Looked-Up are selected randomly.

In the hash implementation, we used the following algorithm for hash function: (a) Multiply a and b as the values of a member in the set, stored into two sequential words, giving c (two-word product); (b) add together the two words of c, giving one value d; (c) Square d, giving e; and (d) Extract the centre 32 bits of e, giving $I_x$ as the hash index. When we got into conflict, a couple of algorithms, namely Sequential Scan and Budgeting [1], are executed in order. If the first one could not solve the problem, the second one is executed.

At the first stage, we ran the software for each single operation Insert, Delete and LookUp and calculated the required time for the operations (see Figure 4). In this stage, the time required to perform each operation was the same. In the second stage, we write scripts of three operations in various combinations and ran the software. In this stage, we got slightly different times required to perform the scripts. The results of this stage are collected in the Table 4.

Table 4. The time required (Micro Second) to perform the operation scripts

| Row | Operations | Methods | | | | |
|---|---|---|---|---|---|---|
| | | Hash | Binary Tree | Link List | Static Array | Our Method (Set) |
| 1 | Insert-LookUp-Delete (ILD) | 0.09 | 0.1 | 0.15 | 0.18 | 0.07 |
| 2 | Insert-Delete-LookUp (IDL) | 0.09 | 0.15 | 0.21 | 0.2 | 0.13 |
| 3 | Lookup-Insert-Delete (LID) | 0.09 | 0.14 | 0.17 | 0.17 | 0.14 |
| 4 | LookUp-Delete-Insert (LDI) | 0.12 | 0.2 | 0.18 | 0.19 | 0.08 |
| 5 | Delete-Insert-LookUp (DIL) | 0.09 | 0.17 | 0.12 | 0.18 | 0.08 |
| 6 | Delete-LookUp-Insert (DLI) | 0.1 | 0.2 | 0.22 | 0.18 | 0.07 |
| 7 | Insert-Insert-Insert (III) | 0.09 | 0.1 | 0.14 | 0.11 | 0.07 |
| 8 | Insert-Insert-LookUp (IIL) | 0.14 | 0.18 | 0.15 | 0.15 | 0.11 |
| 9 | Insert-Insert-Delete (IID) | 0.08 | 0.1 | 0.12 | 0.17 | 0.09 |
| 10 | Delete-Delete-Delete (DDD) | 0.11 | 0.19 | 0.14 | 0.23 | 0.08 |
| 11 | Delete-Delete-Insert (DDI) | 0.1 | 0.12 | 0.15 | 0.17 | 0.12 |
| 12 | Delete-Delete-LookUp (DDL) | 0.09 | 0.12 | 0.15 | 0.23 | 0.1 |
| 13 | LookUp-LookUp-LookUp (LLL) | 0.12 | 0.19 | 0.13 | 0.15 | 0.06 |
| 14 | Lookup-LookUp-Delete (LLD) | 0.07 | 0.1 | 0.19 | 0.1 | 0.09 |
| 15 | LookUp-LookUp-Insert (LLI) | 0.06 | 0.18 | 0.18 | 0.14 | 0.07 |
| | The Average Time | 0.096 | 0.149 | 0.160 | 0.170 | 0.091 |

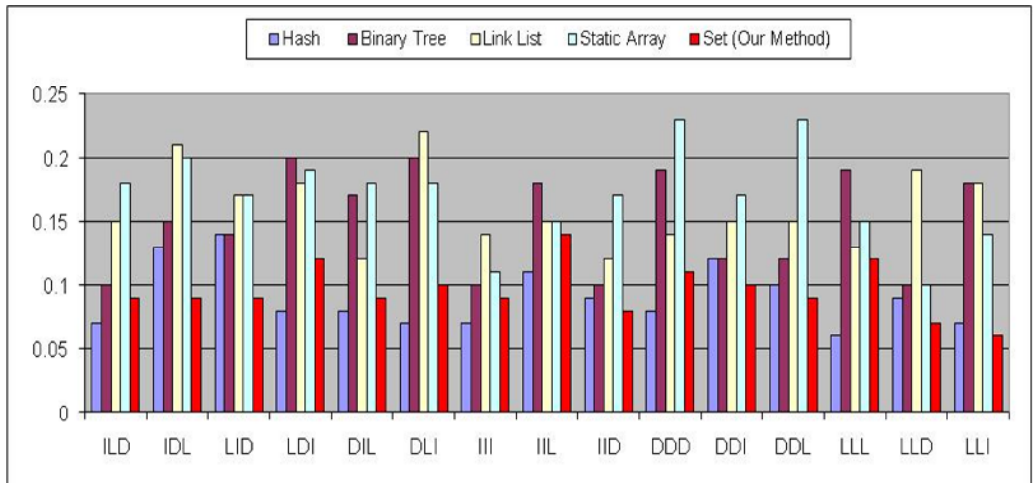

Figure 5. A comparison of the time required to perform the operations

From the information in the Table-4 and Figure-5, we got the following observations:

- **Observation-1**: The average time required to perform the operations in Static Array and Link are longer than that of the others. These methods have approximately the same average time. This observation is consistent with the results in Table 3.

- **Observation-2:** After the Link List and Static Array methods, the Binary Tree has the smallest average time required to do the operations. The average time required to perform the operations in Hash Implementation is shorter than that of the Binary Tree, but it is slightly shorter than that of the Set method.
- **Observation-3:** The average time required to perform the operations in Hash and Set methods is almost the same. The statistical F-test done shows that the differences between our method and others are not significant in the mean value. We believe that the overhead for handling conflict operations was not significant in the operations.

## 5 SUMMARY AND CONCLUSION

Basic differences among the languages refer to the types of data allowed, in the types operations available, and in the mechanisms provided for the implementation. Modern high programming languages need some data structures and their variations in both static and dynamic aspects. In this paper, the standards methods for implementation Property List as Static Array, Link List, Tree and Hash are reviewed. Then a method to implement the property list as Set was presented. The method proposed has more efficiency than the existing methods. In the construction of large application programs, the programmer is almost inevitably concerned with the design and implementation of new data types. The method presented in this paper for implementation of Property List, supports the users so that they can have more flexible and effective dynamic data objects. It can be used in both, at programmer and programming languages levels. For further research, some fast methods could be invented so that some values (names) in a property list have more than one name (value).